\documentclass[fleqn,usenatbib]{mnras}

\usepackage{newtxtext,newtxmath}
\usepackage[T1]{fontenc}

\DeclareRobustCommand{\VAN}[3]{#2}
\let\VANthebibliography\thebibliography
\def\thebibliography{\DeclareRobustCommand{\VAN}[3]{##3}\VANthebibliography}

\usepackage{graphicx}	% Including figure files
\usepackage{amsmath}	% Advanced maths commands
\usepackage{pdflscape} 
\usepackage{subfigure}
\usepackage{multirow}

\title[KNTraP optical search for kilonovae]{An Optically Led Search for Kilonovae to z$\sim$0.3 with the Kilonova and Transients Program (KNTraP)}

\author[N. Van Bemmel, J. Zhang et al.]{Natasha Van Bemmel$^{1,2}$ \& Jielai Zhang$^{1}$,\thanks{E-mail: nvanbemmel@swin.edu.au}
Jeff Cooke$^{1,2,3}$, 
Armin Rest$^{4,5}$, 
Anais M\"{o}ller$^{1,2}$, 
\newauthor
Igor Andreoni$^{2,6,7,8,9}$,
Katie Auchettl$^{10,11}$, 
Dougal Dobie$^{1,2}$, 
Bruce Gendre$^{12,13}$,
Simon Goode$^{2,14}$, 
\newauthor
James Freeburn$^{1,2}$, 
David O. Jones$^{15}$,
Charles D. Kilpatrick$^{16}$,
Amy Lien$^{17}$,
Arne Rau$^{18}$, 
\newauthor
Lee Spitler$^{19,20}$,
Mark Suhr$^{1,3}$,
Fransisco Valdes$^{21}$
\\
$^{1}$Centre for Astrophysics and Supercomputing, Swinburne University of Technology, Victoria 3122, Australia \\
$^{2}$ARC Centre of Excellence for Gravitational Wave Discovery (OzGrav), Victoria 3122, Australia \\
$^{3}$Australian Research Council Centre of Excellence for All-sky Astrophysics in 3 Dimensions (ASTRO 3D)\\
$^{4}$ Space Telescope Science Institute, Baltimore, MD 21218, USA\\
$^{5}$ Department of Physics and Astronomy, Johns Hopkins University, Baltimore, MD 21218, USA\\
$^{6}$Joint Space-Science Institute, University of Maryland, College Park, MD, USA\\
$^{7}$Department of Physics and Astronomy, University of North Carolina at Chapel Hill, Chapel Hill, NC 27599-3255, USA\\
$^{8}$Department of Astronomy, University of Maryland, College Park, MD, USA\\
$^{9}$Astrophysics Science Division, NASA Goddard Space Flight Center, Greenbelt, MD, USA\\
$^{10}$OzGrav, School of Physics, The University of Melbourne, Victoria 3010, Australia\\
$^{11} $Department of Astronomy and Astrophysics, University of California, Santa Cruz, CA 95064, USA\\
$^{12}$The University of Western Australia, OzGrav ARC Centre of Excellence, 35 Stirling Highway, 6009 Crawley, WA, Australia \\ 
$^{13}$University of the Virgin Islands, College of Science and Mathematics, 2 John Brewer's Bay, St Thomas, 00802, VI, USA\\
$^{14}$School of Physics and Astronomy, Monash University, VIC 3800, Australia\\
$^{15}$Institute for Astronomy, University of Hawai'i, 640 N.~A'ohoku Pl., Hilo, HI 96720, USA\\
$^{16}$Center for Interdisciplinary Exploration and Research in Astrophysics (CIERA) and Department of Physics and Astronomy, Northwestern University,\\ Evanston, IL 60208, USA\\
$^{17}$Department of Physics and Astronomy, University of Tampa, 401 W. Kennedy Blvd, Tampa, FL 33606, USA\\
$^{18}$Max-Planck-Institut für Extraterrestrische Physik, Gießenbachstraße, 85748 Garching, Germany\\
$^{19}$Macquarie University Astrophysics and Space Technologies Research Centre, Macquarie University, Sydney, NSW 2109, Australia\\
$^{20}$Australian Astronomical Optics, Faculty of Science and Engineering, Macquarie University, Macquarie Park, NSW 2113, Australia\\
$^{21}$NSF’s National Optical-Infrared Research Laboratory (NOIRLab), 950 N. Cherry Ave, Tucson, AZ 85732, USA\\
}

\date{Accepted XXX. Received YYY; in original form ZZZ}

\pubyear{2024}

\begin{document}
\label{firstpage}
\pagerange{\pageref{firstpage}--\pageref{lastpage}}
\maketitle

\begin{abstract}
Compact binary mergers detectable in gravitational waves can be accompanied by a kilonova, an electromagnetic transient powered by radioactive decay of newly synthesised r-process elements. A few kilonova candidates have been observed during short gamma-ray burst follow-up, and one found associated with a gravitational wave detection, GW170817. However, robust kilonova candidates are yet to be found in un-triggered, wide-field optical surveys, that is, a search not requiring an initial gravitational wave or gamma-ray burst trigger. Here we present the first observing run for the Kilonova and Transients Program (KNTraP) using the Dark Energy Camera. The first KNTraP run ran for 11 nights, covering 31 fields at a nightly cadence in two filters. The program can detect transients beyond the LIGO/Virgo/KAGRA horizon, be agnostic to the merger orientation, avoid the Sun and/or Galactic plane, and produces high cadence multi-wavelength light curves. The data were processed nightly in real-time for rapid identification of transient candidates, allowing for follow-up of interesting candidates before they faded away. Three fast-rising candidates were identified in real-time, however none had the characteristics of the kilonova AT2017gfo associated with GW170817 or with the expected evolution for kilonovae from our fade-rate models. After the run, the data were reprocessed, then subjected to stringent filtering and model fitting to search for kilonovae offline. Multiple KNTraP runs (3+) are expected to detect kilonovae via this optical-only search method. No kilonovae were detected in this first KNTraP run using our selection criteria, constraining the KN rate to $R < 1.8\times10^{5}$ Gpc$^{-3}$ yr$^{-1}$.

\end{abstract}


%%%%%%%%%%%%%%%%%%%%%%%%%%%%%%%%%%%%%%%%%%%%%%%%%%

%%%%%%%%%%%%%%%%% BODY OF PAPER %%%%%%%%%%%%%%%%%%

\section{Introduction}

Kilonovae (KNe) are the thermal emission outburst caused by the merger of a binary neutron star (BNS) or neutron star-black hole (NSBH) system. The optical emission is powered by r-process nucleosynthesis, which is thought to be the process where a large fraction of the heaviest elements in our Universe originate. The extreme nature of these events also allows us to probe the equation of state of dense matter \citep{Bauswein:2017, Margalit:2017, Coughlin:2018, Radice:2018, Nicholl:2021}, and study the extreme physics present as they are associated with gravitational waves (GWs) and gamma ray bursts (GRBs). KNe may also provide a method to independently measure the expansion rate of the Universe \citep{Abbott:2017, Dietrich:2020, Coughlin:2020, Dhawan:2020}.

Despite the great scientific value of KNe, there is only one KN associated with a GW event that has been confirmed to a high degree of certainty electromagnetically and spectroscopically. This KN was discovered as an electromagnetic counterpart to GW170817 \citep{Andreoni:2017, Arcavi2017GW, Chornock:2017, Coulter2017, Cowperthwaite:2017, Drout:2017, Diaz:2017, Evans:2017, Hu:2017, Kasliwal:2017, Kilpatrick:2017, Lipunov2017, McCully:2017, Moller:2017, Nicholl:2017, Shappee:2017, Soares-Santos2017, Pian:2017, Smartt:2017, Tanvir2017, Valenti2017, Utsumi:2017}, a BNS GW event detected during the LIGO/Virgo O2 run in 2017 \citep{Abbott2017GW170817discovery}. The optical counterpart, labelled AT2017gfo, was observed $\sim$11 hours after the time of neutron star merger and had worldwide coverage across all wavelengths. AT2017gfo was a `blue' KN dominated by a lanthanide-poor component and, due to its close proximity to Earth, was observable with optical wavelengths up to $\lesssim$15 days and well fitted to a spherical three component KN model \citep{Villar:2017}. The light curve peaked in the ultraviolet $\lesssim$1 day before quickly fading, followed by a near-infrared peak after a few days.

Compact object mergers have long been proposed to be associated with short-duration GRBs \citep[short GRBs,][]{Blinnikov:1984}, defined as lasting typically less than 2 seconds \citep{kouveliotou93}, and confirmed after observing a short GRB associated with GW170817.
There have been several other KN candidate detections, found as afterglows of short GRBs \citep{Perley:2009, Tanvir:2013, Berger:2013, Gao:2015, Jin:2015, Jin:2016, Jin:2020, Troja:2018, Lamb:2019, Fong:2021, O'Connor:2021, Rossi:2020, Fong:2021, Rastinejad:2021}. However recently, there have also been KN candidates discovered as counterparts to long-duration GRBs (long GRBs) with durations longer and 2 seconds; GRB 060614 \citep{Mangano:2007, Xu:2009, Yang:2015}, GRB 211211A \citep{Rastinejad:2022, Yang:2022, Mei:2022, Gompertz:2023} and GRB 230307A \citep{Levan:2023, Sun:2023, Yang:2024, Gillanders:2023, Dichiara:2023}. See \citet{Nugent:2024} for a comprehensive overview of these and other KN candidates associated with GRBs. 

The rate of KNe, as well as the rates of BNS and NSBH mergers, still have large uncertainties \citep{Mandel:2022}. Short GRB observations suggest a KN rate of $R=270^{+1580}_{-180}$ Gpc$^{-3}$yr$^{-1}$ \citep{Fong:2015} and, for short GRBs based on GW170817A, KN rates have been calculated to be $R=352^{+810}_{-281}$ Gpc$^{-3}$yr$^{-1}$ \citep{DellaValle:2018}, $R=190^{+440}_{-160}$ Gpc$^{-3}$yr$^{-1}$ \citep{Zhang:2018}, and $R=160^{+200}_{-100}$ Gpc$^{-3}$yr$^{-1}$ \citep[SWIFT discoverable,][]{Dichiara:2020}. These rates are however affected by the fact that some some short GRBs have a collapsar origin \citep{Ahumada2021} and some long GRBs are likely produced by BNS mergers \citep[e.g.,][]{Rastinejad:2022, Levan:2023}. Finally, gravitational wave observations currently suggest the intrinsic rates for BNS and NSBH mergers to be $R_{\textrm{BNS}} = 210^{+240}_{-120}$\,Gpc$^{-3}$\,yr$^{-1}$ and $R_{\textrm{NSBH}} = 8.6^{+9.7}_{-5.0}$\,Gpc$^{-3}$\,yr$^{-1}$ \citep{theligo-virgo-kagracollaboration_observing_2023, coughlin_ligo/virgo/kagra_2022, singer_lpsinger/observing-scenarios-simulations:_2022}. 

Optical wide-field surveys are yet to produce viable KN candidates \citep[but see][]{McBrien2021}. Previous searches constrained the KN rates assuming a variety of models, given that the KN luminosity function is still unknown. After GW170817, optical surveys could constrain the KN rate using AT2017gfo as a reference. Using DLT40, \cite{Yang:2017} calculated a upper limit of $R < 9.9\times10^{4}$ Gpc$^{-3}$ yr$^{-1}$. \cite{Smartt:2017} calculated an upper limit to the KN rate using a non-triggered ATLAS survey, to be $R < 3.0\times10^{4}$ Gpc$^{-3}$ yr$^{-1}$. A search conducted using the first 23 months of data from the Zwicky Transient Facility \citep[ZTF,][]{Bellm:2019, Graham:2019} constrained the GW170817-like KN rate to $R < 1775$ Gpc$^{-3}$ yr$^{-1}$ \citep{Andreoni:2020}. The program was expanded with a real-time analysis component \citep[ZTFResT, ][]{Andreoni:2021} which improved this constraint further to $R < 900$ Gpc$^{-3}$ yr$^{-1}$. The most constraining optical rate is from \cite{Kasliwal:2017}, who find an upper limit of $R < 800$ Gpc$^{-3}$ yr$^{-1}$, by conducting a galaxy-targeted search strategy via catalog cross-matching with the Palomar Transient Factory (PTF). 

To optimise our chances of discovering more KNe and provide further constraints on the true KN rate, we cannot rely purely on GW and GRB alerts, as they are only sensitive to certain merger system orientations, whereas most other wavelength emission is isotropic \citep{Metzger:2017}. In addition, GW and GRB sky uncertainty regions are often large, making covering the regions with KN searches difficult before the events fade, they can include regions that are near the Sun, making coverage with many other wavelengths difficult, or they can cross the Galactic plane, where the high extinction can make certain wavelength detection difficult. Many GRBs are too distant for associated KN detection or, if detectable, have a GRB afterglow present which can contaminate/overpower the KN signature. Furthermore, GW detectors are not online all the time and when online, KN searches via this method are limited to their detection horizon. Currently in O4, the LIGO/Virgo/KAGRA (LVK) detectors have a orientation-averaged range of 190 Mpc, and horizon of 430 Mpc \citep{Abbott:2020}. 

Finding KNe independent of GRB and GW triggers via `blind' day-cadenced optical searches also allows us to sample the light curve at early times, while the time lost in finding the transient with a `reactive' follow-up search, given the GW/ GRB localisation area is avoided. Instead of a reactive search, blind optical searches continuously sample the sky and randomly sample KN light curves in a `proactive' manner. This makes it possible to detect KNe very early in some cases, possibly earlier than can be done with reactive approaches (depending on their localisation areas and source brightnesses). AT2017gfo was found relatively quickly, in $\sim$11 hours, however it was a particularly close event at only $\sim$40 Mpc, and future GW follow-up may not be as fast or comprehensive, while potential KN candidates may not be as close. While optical follow-up of GW detections may lose some light curve information, for optically led KN detections, GW archival data can be searched for the event after the fact, and searched for low signal-to-noise (SNR) events (given the detectors were online) to include GW information. 

Characterising KN light curves at early times can provide valuable information to help constrain theoretical models \citep[e.g.,][]{Drout:2017, Kasliwal:2017, Arcavi:2018, Piro_Kollmeier:2018, Waxman:2018}, and early sampling can inform us on their physical environments, heating and cooling mechanisms, and ejecta structure. To perform the science required to characterise the distribution of these KNe properties, investigate the neutron star equation of state, and measure the expansion rate of the Universe, a much larger population of KNe that what exists is required. The current output of GW and GRB searches ($\sim$3 KNe per 7 years, only one of which associated with a GW and spectroscopically confirmed) is insufficient to satisfy the required number of KNe detections to do this science in a practical timeframe. To increase the discovery rate, we present a blind optical survey called the Kilonova and Transients Program (KNTraP). This survey provides an efficient search for KN candidates with the advantages over conventional GW and GRB follow-up searches as presented above. 

The paper is organized as follows. The motivations underlying the KNTraP design are outlined in Section \S\ref{sec:kntrap_motivation}. We describe the observations and data processing in Section \S\ref{sec:data}and the KN models used in the candidate analysis in Section \S\ref{sec:kn_models}. The candidate selection is covered in section \S\ref{sec:analysis}, and the results, including the most interesting candidates from the real-time and offline pipelines are presented in \S\ref{sec:results}. Section \S\ref{sec:rates} describes constraints on KN rates by our data. Finally, a discussion of the results are presented in Section \S\ref{sec:discussion} and summarised in Section \S\ref{sec:summary}.

\section{KNTraP design}
\label{sec:kntrap_motivation}

A deep, wide-field, nightly cadenced optical search would overcome the limitations of current GRB and GW follow-up searches and probe much deeper into the Universe. To maximise the efficiency and volume probed by such a survey, the depth, area on sky, and cadence must be balanced. The number of filters used must also be considered, as colour information is vital to help differentiate KN candidates from other fast-evolving transients. Minimising the number of necessary filters helps maximise the area and depth for a given time on sky. Classically scheduling such a survey (i.e., no Target of Opportunity disruptions) is easier for observatories and avoids impact on other scientific programs. In addition, a classically scheduled program that also has quick data processing would allow better organisation of follow-up facilities for larger wavelength coverage and spectroscopy to help classify any KN candidates. KNTraP was designed to provide such a deep survey with a wide field of view and a nightly cadence. 

One main aim of KNTraP is to probe the maximum cosmological volume during one observing night to then be repeated consecutive nights. As such, a 4m- to 8m-class optical telescope with a wide-field imager (here, defined as $>$1 deg$^2$ field of view, FOV) is needed. The only wide-field imagers on 4m-class and larger telescopes are the Dark Energy Camera \citep[DECam, 3 deg$^{2}$ FOV,][]{Flaugher:2015} on the 4m Blanco telescope at the Cerro Tololo Inter-American Observatory and the Hyper Suprime-Cam imager \citep[HSC, 1.8 deg$^2$ FOV,][]{miyazaki18} on the 8m NAOJ Subaru telescope. For this program, Subaru HSC is not well suited, as the time required for filter change makes multiple filter coverage of the many targeted fields per night less feasible. Although DECam reaches $\sim$1 mag shallower per given exposure time than HSC, it covers more area per pointing and it has a fast filter change (within the $\sim$30\,s readout time). 

In order to distinguish a KN from other more prevalent types of optical transients, such as supernovae or stellar flares, imaging in the \emph{g} and \emph{i} filters was adpoted. As shown by \citet{Andreoni:2019}, \emph{g}-\emph{z} or \emph{g}-\emph{i} colour difference provides a large lever arm in colour, which allows us to identify the fast colour evolution expected for KNe. While DECam is more sensitive towards the infrared than many other optical cameras, its throughput in the \emph{z}-band is sufficiently lower than the \emph{i}-band, which would significantly lower the overall volume probed with the same observing time. Therefore, the \emph{i} band filter was deemed the optimal redder filter choice as it requires significantly less exposure time than the \emph{z} band. 

\begin{figure*}
    \centering
    \includegraphics[width=1\linewidth]{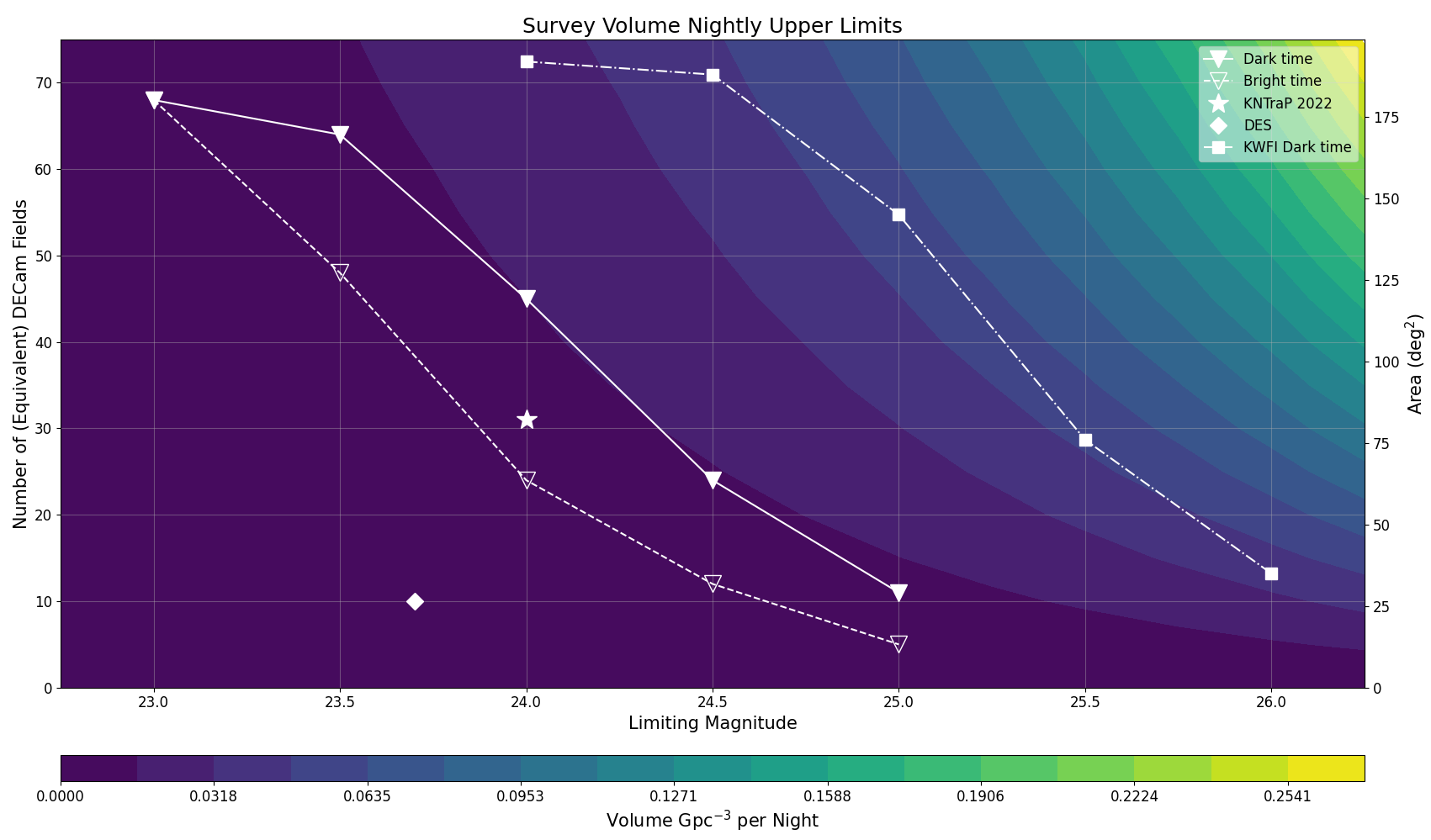}
    \caption{The volumes which can be achieved with different depths and number of fields (area on sky) observed per night, with overheads considered (i.e. exposure time, filter exchange, readout, and dithering). The volume is calculated using a DECam FOV, to a distance where the AT2017gfo peak magnitude can be detected. Inverted triangles show the upper limits that can be reached for DECam in dark (solid) and bright (hollow) time. The volume from this KNTraP run is shown as a star, and is above the bright time limit as we had a combination of deep and shallow fields. The DES nightly volume is shown here for comparison \citep{Smith:2020}, depth averaged between their deep and shallow fields, as well as the projected KWFI limits for its current design.}
    \label{fig:volume_limits}
\end{figure*}

To maximise the sky volume probed, a balance was struck between number of fields ($\sim$30 fields, $\sim$100 deg$^{2}$) targeted each night and depth (m $\sim$ 24--25) in multiple filters for the consecutive nights during the observing run. Figure \ref{fig:volume_limits} demonstrates the trade-off for depth and number of fields for DECam. Comparisons between other surveys and different observing conditions are shown on this plot also. 

KNTraP data were processed nightly for next day candidate vetting. This day-scale cadence and multi-filter data allowed us to distinguish KNe from other variable or transient sources using the criteria discussed in this work. For promising KN candidates during this first KNTraP observing run, a number of follow-up observations were coordinated using a subset of the Deeper Wider Faster \citep[DWF,][]{Andreoni_Cooke:2017} facilities in multiple wavelength regimes, as well as public ToO request opportunities (\S\ref{ssec:realtime-candidates}).

Assuming an AT2017gfo-like event with a peak magnitude of M $=-16.7$ in the \emph{i}-band and adopting the BNS rate constrained by LVK at the time of survey design, $320^{+490}_{-240}$Gpc$^{-3}$yr$-1$ \citep{Abbott:2021}, we estimate on average $\sim$0.3 KNe to be discovered per 11 day observing run with the KNTraP observing strategy and an expected limiting magnitude of m $\sim$ 24.5 mag in \emph{g} and \emph{i}-bands performed in bright/grey time for this first run. The magnitude limit gives a KN discovery magnitude of m $\sim$ 23.5, to allow for $\geq$ 1 mag of light curve information. 

Finally, KNTraP is sensitive to other fast evolving transients which have not been previously well studied with traditional longer cadence, single filter, or shallower surveys. Examples of these are fast-evolving luminous blue optical transients (FBOTs) \citep{Drout:2014, Arcavi:2016, Tanaka:2016, Pursiainen:2018, Prentice:2018, Rest:2018, Ho:2023}, the early evolution of Type Ia supernovae \citep{Cao:2015, Miller:2020}, GRB "orphan" afterglows \citep{Dalal:2002, Huang:2002, Nakar:2002, Rhoads:2003}, core collapse supernova shock breakouts \citep{Waxman:2017, Soderberg:2008}, tidal disruption events \citep[][and references therein]{Andreoni:2022, Gezari:2021}, stellar flares \citep[][and references therein]{Kowalski:2024}, novae \citep{Strope:2010, Schaefer:2010}, and cataclysmic variables \citep{Gansicke:2009, Pala:2020}.

\section{Data}
\label{sec:data}

The target selection criteria is described in section \S\ref{ssec:data_aquisition}. Section \S\ref{ssec:data_processing} details how the data was processed in real-time, and briefly how candidates were identified in both the real-time and offline pipelines.

\subsection{Data Acquisition}
\label{ssec:data_aquisition}

\begin{figure}
    \centering
    \begin{subfigure}
        \centering
        \includegraphics[width=0.49\textwidth]{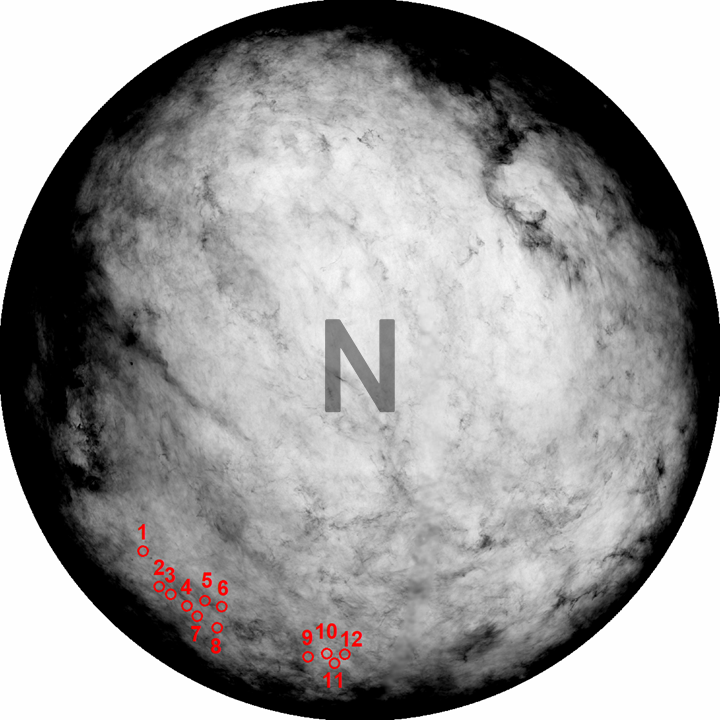}
    \end{subfigure}
    \hfill
    
    \begin{subfigure}
        \centering
        \includegraphics[width=0.49\textwidth]{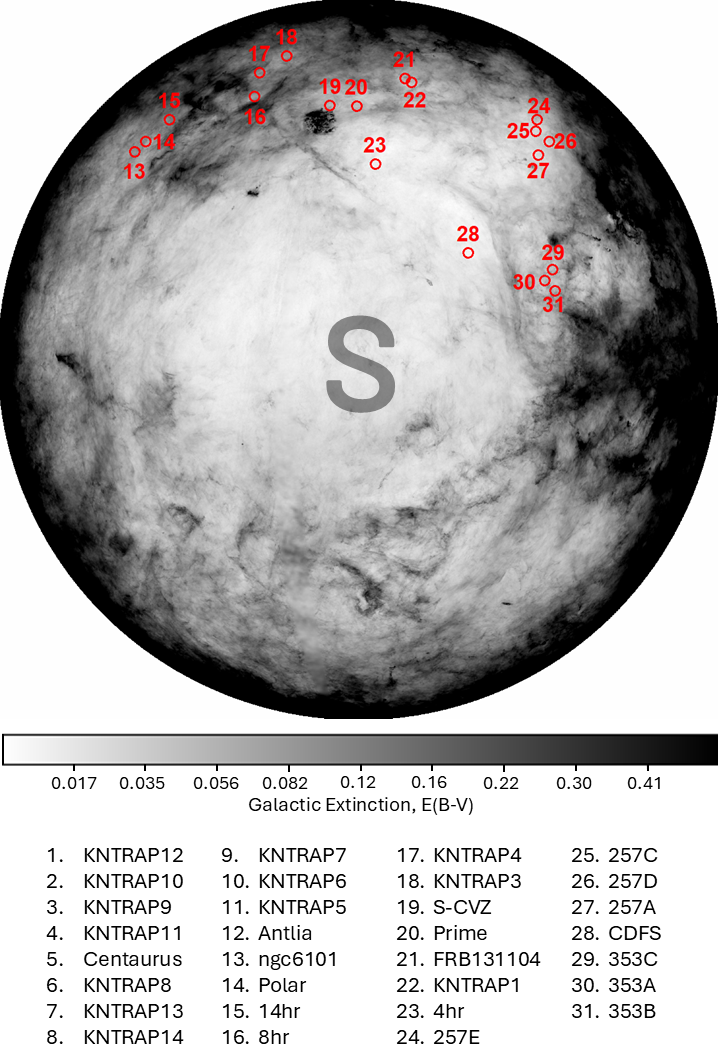}   
    \end{subfigure}
    
    \centering
    \caption{Dust maps of the Northern and Southern hemispheres \citep[][SFD]{SFD:1998} with KNTraP fields overlaid (red circles, numbered). The sizes of the red circles are indicative of the FOV for each field. The Galactic plane is along the edge of both images, and Galactic poles in the centres. Dark (light) regions denote areas of high (low) extinction.}
    \label{fig:field_locations}
\end{figure}

We conducted observations using DECam over 11 nights from February 12 to 22 in 2022. These observations targeted 31 fields in \emph{g}- and \emph{i}-band filters, with 3x140 (3x60) and 3x170 (3x70) second exposures, respectively, for our target fields termed deep (shallow) fields. The fields were specifically chosen to minimise Galactic extinction given the sky visibility at the time of observations with a threshold of $E(B-V) <$  0.1. The dates of the scheduled run required us to search for low Galactic extinction regions closer to the plane of the Galaxy than originally planned. Fields ideally targeted known galaxies and areas with high galaxy density (within z $<$ 0.25, 1000 Mpc) from the GLADE catalog. Fields were selected to avoid the Moon as observations took place through bright/grey time, and pointings adjusted to avoid bright stars. Targets were also prioritised if they had preexisting DECam imaging, and were roughly equatorial to maximise the number of facilities which can perform follow-up observations. Table \ref{tab:kntrap_fields} gives a list of fields observed, and Figure \ref{fig:field_locations} shows the locations of the fields overlaid on a dust map of the sky. 

Ten of the 31 fields have been previously observed by the DWF program; CDFS, 4hr-field, Prime, FRB131104, 8hr-field, Antlia, Centaurus, 14hr, Polar, and NGC6101. Seven fields are pointings observed by the Young Supernova Experiment \citep[YSE,][]{Jones:2021}; 257A, 257D, 257C, 257E, 353A, 353B, and 353C. One field is the Southern Continuous Viewing Zone (SCVZ), which is a field in a location which can be observed without break by space telescopes (e.g., TESS). The remaining are new pointings for KNTraP which met the criteria listed above. 

\begin{table}
    \centering
    \caption{KNTraP target fields, central J2000 coordinates and the Galactic extinction at the field centre. Shallow fields had a single exposure time of 70\,s (60\,s) in the \emph{i} (\emph{g}) filter per night and deep fields had three exposures of 170s (140s) in the \emph{i} (\emph{g}) filter per night.
    }
    \label{tab:kntrap_fields}
    \begin{tabular}{c|c|c|c|}
        \hline
        Field& RA& Dec& E(B-V)\\
        \hline
        \multicolumn{4}{|c|}{\emph{shallow fields}}\\
        \hline
         353A & 60.02 & -11.40 &  0.051\\
         353B & 60.02 & -8.30 &  0.063\\
         353C & 63.21 & -11.40 &  0.054\\
         257A & 85.67 & -24.29 &  0.028\\
         SCVZ & 90.00 & -66.56 &  0.064 \\
         \hline
         \multicolumn{4}{|c|}{\emph{deep fields}}\\
         \hline
         CDFS & 52.50 & -28.10 &  0.007\\
         4hr & 62.50 & -55.00 &  0.014\\
         Prime & 88.78 & -61.35 &  0.045\\
         257D & 90.57 & -22.69 &  0.038\\
         257C & 91.57 & -25.79 &  0.030\\
         257E & 94.97 & -25.79 &  0.050\\
         FRB131104 & 101.00 & -51.27 &  0.070\\
         KNTRAP1 &  99.20 & -50.20 &  0.049\\
         8hr &  124.00 & -78.75 &  0.086\\
         KNTRAP3 & 133.50 & -67.80 &  0.090\\
         KNTRAP4 & 138.50 & -73.60 &  0.131\\
         KNTRAP5 & 157.50 & -38.50 &  0.083\\
         Antlia & 157.50 & -35.33 &  0.102\\
         KNTRAP6 & 160.75 & -37.30 &  0.055\\
         KNTRAP7 & 163.50 & -40.30 & 0.081\\
         KNTRAP8 & 188.40 & -40.30 &  0.073\\
         Centaurus & 192.50 & -41.33 &  0.120\\
         KNTRAP9 & 200.90 & -45.15 &  0.099\\
         KNTRAP10 &  204.40 & -45.15 &  0.094\\
         14hr &  218.50 & -78.10 &  0.122\\
         Polar & 240.00 & -75.00 &  0.099\\
         NGC6101 & 246.50 & -73.00 &  0.087\\
         KNTRAP11 & 195.75 & -45.30 &  0.084\\
         KNTRAP12 & 212.00 & -39.80 &  0.072\\
         KNTRAP13 & 192.00 & -46.10 &  0.094\\
         KNTRAP14 & 186.12 & -45.90 & 0.084\\
         \hline
    \end{tabular}
\end{table}

\begin{table}
    \centering
    \caption{Observing conditions for each night. Seeing was averaged over each night. Lunation is the illumination percentage. Angular separations are between the Moon and field centres.}
    \label{tab:kntrap_weather}
    \begin{tabular}{c|c|c|c|c}
        \hline
         MJD& Average Seeing& Lunation& \multicolumn{2}{c}{Angular Separation (deg)} \\
         & & &  Minimum & Average \\ 
         \hline
         59622& 1.14& 72.5\% & 46.65 & 89.03\\
         59623& 1.35& 80.5\% & 51.61 & 87.49\\
         59624& 1.25& 87.5\% & 55.66 & 86.09\\
         59625& 1.07& 93.1\% & 60.94 & 84.76\\
         59626& 1.21& 97.2\% & 56.59 & 83.61\\
         59627& 1.12& 99.6\% & 50.83 & 82.66\\
         59628& 1.19& 99.9\% & 47.12 & 82.04\\
         59629& 1.03& 98.1\% & 46.09 & 81.89\\
         59630& 0.87& 94.1\% & 39.96 & 82.24\\
         59631& 1.09& 88.1\% & 36.15 & 83.27\\
         59632& 1.29& 80.1\% & 34.76 & 85.00\\
         \hline
    \end{tabular}
\end{table}

\subsection{Data Processing}
\label{ssec:data_processing}

The data collected were processed rapidly in real-time as described in Section \S\ref{sssec:realtime_processing}, and processed again later offline with more carefully selected parameters detailed in Section \S\ref{sssec:offline_processing}. In real-time, image subtraction was performed as quickly as possible, and all candidates meeting a basic transient criteria were passed to visual inspection to be conducted nightly. The offline selection pipeline was designed after the observing run to reprocess the data thoroughly and search for convincing AT2017gfo-like and other KN candidates. The offline pipeline had to be more conservative than the real-time search, as there was no longer the ability to follow-up candidates. The candidate selection criteria of the offline filtering pipeline is described in further detail in \S\ref{sec:analysis}.

\subsubsection{Real-Time Processing}
\label{sssec:realtime_processing}

The DECam raw images were submitted to the NOIRLab Community Pipeline \citep{Valdes:2014} in batches as the data were collected off the telescope. Once the NOIRLab procerssing was complete, the data were transferred to the OzSTAR supercomputer at the Swinburne University of Technology and at the end of the night, passed through the {\tt Photpipe} \citep{Rest:2005,Rest:2014} image subtraction and candidate identification pipeline. {\tt Photpipe} performed image calibrations, astrometric calibrations, warping and coaddition of the images from the telescope using {\tt SWarp} \citep{Bertin:2002}. A difference image was produced from the subtraction of the science and template images with {\tt HOTPANTS} \citep{Alard:2000, Becker:2015}. Photometry was done on the difference images using {\tt DoPhot} \citep{Schechter:1993}, to prepare the images for  candidate identification. 

For majority of the fields, the stacked images from the first night's data were used as the template image for deep fields, and a single exposure for shallow fields. It is preferred to have archival data to use as a template, but these did not exist for several newly targeted fields and were not imaged at an adequate depth for other fields. For most shallow fields (257A, 353A, 353B, 353C), there were adequate template images from archival data, which were used. 

Transient candidates were produced based on a SNR of the source and the number of times it was detected. For our candidate selection, we used a SNR > 3$\sigma$ and a minimum of 3 detections. Photometry was measured for the sources where a detection in the difference image was present, and forced photometry performed at the location of the source at all other nights even if the source was not detected. Thumbnail images of the transient candidates were generated, as well as photometric light curve plots, and presented on an online web interface for nightly visual inspection.  

The process from data collection to the beginning of visual inspection took less than 24 hours. The candidates included artefacts and poor subtractions and not all could be visually inspected each night due to the small team. As only the the {\tt Photpipe} filtering was applied, many contaminants polluted these candidate lists which could be removed with a more strict filtering pipeline. This filtering would significantly increase the speed of human vetting.

\subsubsection{Offline Processing}
\label{sssec:offline_processing}

In the post-run offline processing, {\tt Photpipe} was used again on the data. The convolving parameters were adjusted in this reprocessing, to improve the quality of the difference images. In real-time, {\tt HOTPANTS} was set up to always convolve the template image to the science before performing the image subtraction. Offline, this was changed so that the convolution would occur on either the template or science image depending on which was "better", based on PSF widths. This was important, as the candidate photometry is performed on the difference images. In addition, the number of detections required for candidate detection was decreased to require a minimum of two detections. This criteria was loosened as more stringent filtering was applied after this stage.

\section{Generated AT2017gfo-like Sources in KNTraP}
\label{sec:kn_models}

From the candidates identified offline with {\tt Photpipe}, we first aimed to identify candidates which appear exactly like AT2017gfo. For this purpose, we created a simulated grid of KN light curves using the AT2017gfo light curve as a reference. Although there is a broad distribution of parameters in various KN model light curves, and they can vary from AT2017gfo, for this initial aim 
%only one model is simulated. 
%These simulated models would help quantify the number of AT2017gfo-like KNe found (or missed) by our searches given the cadence and selection criteria that we employed. Some of the selection criteria detailed in \S\ref{ssec:candidate_selection} is informed and validated by these models, to ensure any simulated KN that would be considered a strong candidate is not rejected.
we used the spherical three-component model best fit as computed by \cite{Villar:2017}. This model was shifted by starting date and redshift, and then sampled once nightly in both \emph{i} and \emph{g} filters to simulate all possible KNe that could be observed by KNTraP. Time-dilation was considered, and redshift-related filter corrections were incorporated by shifting \emph{i}-band magnitudes towards \emph{r}-band, and \emph{g}-band towards \emph{u}-band in steps relative to redshift. The observing conditions shown in Table \ref{tab:kntrap_weather} were used to calculate magnitude limits. The limits and magnitude uncertainties were applied to the model data for efficiency calculations.

We considered redshifts of 0.01 ($\sim$40 Mpc, the distance of AT2017gfo) to 0.25 ($\sim$1000 Mpc) in small increments. The upper redshift limit was informed by our magnitude limits. The step sizes were determined such that the \emph{i}-band peak magnitude of the KN light curve from step to step did not have a difference larger than $\sim$0.3 mags. This ensured consistent sampling throughout the simulation grid. The step sizes for the following redshift bins were 0.001 (0.01 $\leq$ \emph{z} < 0.014), 0.002 (0.014 $\leq$ \emph{z} < 0.03), 0.005 (0.030 $\leq$ \emph{z} < 0.06), and 0.01 (0.06 $\leq$ \emph{z} $\leq$ 0.25).

To help account for KNe which may have burst at any time throughout the observing run, simulated KNe were created to burst on the first night, through to the final night, in steps of 8 hours. The decision to have this step size was based on the limitations of the model rise information and the shape of the nightly light curves to capture the main variations, without generating an excessive number of simulated light curves. When calculating the efficiency of recovered simulated KNe, 8 hour steps does not reduce the efficiency compared to steps of six hours. A representative sample of the model grid is shown in Appendix \ref{appsec:model_grid}. The simulations were also a necessary tool to measure KN rates, discussed in Section \S\ref{sec:rates}.

\section{Analysis of Offline Candidates}
\label{sec:analysis}

\subsection{Candidate Selection}
\label{ssec:candidate_selection}

The transient candidates produced from {\tt Photpipe} were passed through a critical multi-step filtering pipeline to identify and assess KN candidates. This filtering pipeline is only valid for the offline processing in Section \S\ref{sssec:offline_processing} and analysis of the data. The selection criteria for the analysis in this paper are as follows.

\subsubsection{Minimum number of detections}
\label{sssec:minimum_detections}

We required $\geq 3$ detections in any band to be present for each candidate. This criterion enabled the rejection of cosmic rays, CCD artefacts, and spurious effects that are likely to appear in the images. Because of the nightly cadence in 2 bands, requiring at least 3 detections instead of two not only decreased the chance coincidence of two spurious detections, but it also made it more likely to have both brightness evolution and color information in the light curves. This was necessary as the {\tt Photpipe} image subtraction pipeline does not have a real/bogus classification system in place, such as those used in other wide-field surveys which employ machine learning techniques (e.g. \cite{Cao:2016} for PTF, \cite{Duev:2019} for ZTF and YSE). This criteria eliminated faint and distant KNe that may only have one or two detections above the limiting magnitude threshold and those that may have burst two or less days before the end of the run.

\subsubsection{Stellarity and known variability}
\label{sssec:stellarity}

We cross-match all the candidates with catalogs to reject likely stars and known variable sources such as AGN and QSOs using the following criteria. Unless specified, we used a conservative cross-match radius of 1.5$''$.

\begin{itemize}
    \item Candidates spatially coincident with sources present in the Gaia DR3 catalog \citep{Gaia2021DR3} with parallax $> 1.081$\,mas were rejected. This threshold for stellarity, employed also in KN searches in ZTF data by \cite{Andreoni:2020}, was determined from the study of AllWISE catalog quasars present in the Gaia data set \citep{Luri2018}
    \item Candidates spatially coincident with sources in the Legacy Survey DR9 catalog, which were well fit by a tractor model with type = ``PSF", were rejected
    \item Candidates spatially coincident with sources in the Million Quasars (Milliquas) catalogue, version 8 \citep{Flesch2023}
    \item Candidates spatially coincident within 2$''$ of sources classified as variables in the SIMBAD catalogue were rejected \citep{Moller:2021}.   
\end{itemize}

\subsubsection{First night detection}

We only want to search for KNe that have exploded during our run, as we cannot constrain the light curve prior to observations. Since for the majority of our target fields, the first night's data was used as the template image, having a non-detection on the first science night was especially important to ensure identification of the fast rise of KNe and because recently burst transients and persistent variable sources are more difficult to distinguish. As a result, candidates are rejected if there is a detection of a PSF-like source on the first night.

\subsubsection{CCD edge proximity}
\label{ccd_edge}

Candidates located near the CCD edges were identified and those within 20 pixels of an edge were rejected. The pixel threshold was based on PSF sizes of objects in the images and the telescope re-pointing accuracy. This procedure helped eliminate possible artefact transients and ensured the transient candidate photometry was not be affected by the CCD edge.

\subsubsection{Light curve properties I: consecutive detections}
\label{sssec:consecutive_detections}

It is expected that a real KN would be detected over multiple consecutive nights in the \emph{g} and \emph{i} filters. The next filtering step counts the number of consecutive nightly detections for each candidate. These detections are also required to be PSF-like, meaning they must meet a certain criteria mainly based on {\tt Source Extractor} \citep{Bertin:1996} parameters: must have values of ellipticity < 0.7, FWHM < 2 times the seeing on the night (information gathered through automated nightly observing log tool), spread model between -0.02 and 0.02, as well as having a SNR > 5.

We allow for one light curve "hole" (one night of missing data) in the consecutive detection count in the event of poor subtraction (perhaps due to poor seeing on a given night), poor alignment, etc. The candidates were required to have $\geq$ 3 consecutive detections (allowing one hole) in the  \emph{i} band. The cut was focused on \emph{i} band because KNe are expected to be red, with some not having a blue component and, thus, more likely to be observed in the \emph{i} band.

\subsection{AT2017gfo-like Candidates}
\label{ssec:chi2_modelfitting}

To identify any candidates with an AT2017gfo-like light curve, a $\chi^2$ goodness of fit test was performed, comparing the candidates which passed the data quality cuts (described in section \S\ref{ssec:candidate_selection}) to the simulated KNe (Section 
\S\ref{sec:kn_models}).  Over 1400 candidate light curves were compared with 1026 KN models from the model grid. Each candidate had differing \emph{g} and \emph{i}-band light curves and various complications to consider. For candidate light curves with missing data points or "holes" in the light curve, the model data point was used to compare to the limiting magnitude. If the model magnitude was deeper than the limiting magnitude, this data point was "masked" and did not contribute to the $\chi^2$ calculation, as observing conditions did not allow for this data to be observed.
However, if the model magnitude was shallower than the limiting magnitude, the limiting magnitude was used at the data point in the calculation. This means that light curves that have missing data points where the model is much brighter than the limiting magnitude, are "penalised".  

Because we do not have light curve information for the rise of AT2017gfo, at $<$ 1 day, and we have data just prior to the peak, we required that the time of the first detection of the candidate and model must match within 24 hours. 

The candidates were ranked from highest to lowest p-value. The candidates with a p-value greater than 0.001 (0.1$\%$ probability) were investigated further individually by visual inspection of their light curves and thumbnail images.

\subsection{Rapid Evolution Candidates}
\label{ssec:rapid_evolution}

As the KNe population is not well sampled, there may be KNe which do not have light curve behaviour identical to AT2017gfo. To find these possible KNe candidates, we apply a general selection cut which search for rapidly evolving candidates which match theoretical expectations of KNe \citep[e.g.,][]{Rosswog:2005, Metzger:2010}. 

%\subsubsection{Light curve properties II: rapid evolution}

To search for candidates that follow the expected evolutionary behaviour of KNe, the rise and fade rates of the candidate light curves are determined by measuring the magnitude changes between detections. The selected limits on KN magnitude evolution rates are based upon a KN model grid analysis by \citet{Andreoni:2020} using POSSIS \citep{Bulla:2019} and the KN model subset grids by \citet{Dietrich:2020} for NSBH and \citet{Anand:2020} for BNS. For NSBH (BNS) the 95\% threshold magnitude fade rates for the KN model grid were 0.57 (0.60) and 0.27 (0.43) mag day$^{-1}$ for  \emph{g} and  \emph{i} band respectively. \citet{Kasliwal:2020} also performed a KN search and used evolution thresholds where any event rising slower than 1 mag day$^{-1}$ or fading slower 0.3 mag day$^{-1}$ would not be considered. 

With the cadence of our survey, we want to consider both rise and fade. We adopt the criteria $<$1 mag day$^{-1}$ rise and $>$ 0.3 mag day$^{-1}$ fade to select candidates. All candidates that pass these rapid evolution cuts have light curve plots and thumbnail images generated for visual inspection. 

The selected candidates were separated qualitatively into different classes; artefact/bad column, bad subtraction, or real source. The real sources were then further classified as either candidates with large variations (changes of $\gtrsim$1 mag over the entire light curve), small variations ($\lesssim$1 mag), or fast fading (changes of $\gtrsim$1 mag, without periodic behaviour present in the light curve). Table \ref{tab:candsummary} shows a summary of the number of candidates remaining after each filtering step.

\begin{landscape}
    \begin{table}
        \centering
        \caption{Initial candidates produced by {\tt Photpipe} and subsequent survival numbers after various selection criteria, see Section \S\ref{sec:analysis} for selection details. The second to last row shows the total candidates for all fields after each step. The final row shows the number of real-time candidates after each filtering step.}
        \label{tab:candsummary}
        \begin{tabular}{c|c|c|c|c|c|c|c}
        \hline
        Field &Initial candidates& Cross-matches removed&$\geq$ 3 detections& First night removed& CCD edge proximity removal& $\geq$3 consecutive detections& Rapid evolution\\
        \hline
        14hr & 15106 & 10015 & 2092 & 2092 & 1977 & 90 & 35 \\
        257A & 9069 & 8614 & 1025 & 703 & 703 & 15 & 10 \\
        257C & 8564 & 8138 & 2101 & 2101 & 2077 & 42 & 21 \\
        257D & 9582 & 9171 & 2906 & 2906 & 2880 & 68 & 32 \\
        257E & 5618 & 5280 & 1286 & 1286 & 1272 & 53 & 33 \\
        353A & 1887 & 1858 & 162 & 120 & 120 & 3 & 0 \\
        353B & 3248 & 3232 & 186 & 145 & 145 & 5 & 1 \\
        353C & 3100 & 3052 & 360 & 292 & 292 & 16 & 1 \\
        8hr & 8081 & 7713 & 1922 & 1922 & 1815 & 32 & 23 \\
        Antlia & 10437 & 10104 & 2858 & 2858 & 2845 & 36 & 18 \\
        CDFS & 15518 & 15256 & 3889 & 3889 & 3789 & 30 & 6 \\
        Centaurus & 19270 & 19024 & 2538 & 2538 & 2522 & 50 & 30 \\
        FOURHR & 4990 & 4773 & 1022 & 1022 & 1016 & 11 & 4 \\
        FRB131104 & 4790 & 4164 & 1082 & 1082 & 1075 & 40 & 24 \\
        KNTRAP10 & 12329 & 12009 & 1874 & 1874 & 1863 & 77 & 37 \\
        KNTRAP11 & 15059 & 14679 & 2431 & 2431 & 2423 & 68 & 44 \\
        KNTRAP12 & 9427 & 9025 & 1406 & 1406 & 1393 & 63 & 35 \\
        KNTRAP13 & 10071 & 9678 & 1746 & 1746 & 1737 & 64 & 37 \\
        KNTRAP14 & 8642 & 8445 & 1695 & 1695 & 1686 & 14 & 10 \\
        KNTRAP3 & 8914 & 8522 & 1721 & 1721 & 1685 & 53 & 40 \\
        KNTRAP4 & 9985 & 9682 & 2242 & 2242 & 2208 & 48 & 23 \\
        KNTRAP5 & 11773 & 11450 & 3426 & 3426 & 3377 & 57 & 25 \\
        KNTRAP6 & 17938 & 17546 & 6015 & 6015 & 5968 & 87 & 47 \\
        KNTRAP7 & 14650 & 14368 & 2740 & 2740 & 2730 & 53 & 25 \\
        KNTRAP8 & 20708 & 20426 & 3945 & 3945 & 3900 & 49 & 32 \\
        KNTRAP9 & 18985 & 18682 & 2892 & 2892 & 2881 & 67 & 47 \\
        KNTRAPaa & 3584 & 3175 & 681 & 681 & 674 & 15 & 10 \\
        NGC6101 & 11949 & 11571 & 1616 & 1616 & 1543 & 91 & 55 \\
        Polar & 13988 & 13677 & 2187 & 2187 & 2134 & 84 & 55 \\
        Prime & 7340 & 6990 & 1916 & 1916 & 1873 & 22 & 8 \\
        SCVZ & 12798 & 12030 & 1337 & 1337 & 1329 & 31 & 8 \\
        \hline
        Total: & 327400 & 312349 & 63299 & 62826 & 61932 & 1434 & 776 \\
        \hline
        Real-Time Candidates & 23 & 23 & 23 & 13 & 13 & 4 & 0 \\
        \hline
        \end{tabular}
    \end{table}
\end{landscape}

\section{Results}
\label{sec:results}

We present the results from the real-time search in \S\ref{ssec:realtime-candidates}, including the details of the most interesting candidates found. The results from the offline pipeline search are presented in \S\ref{ssec:offline_candidates}.

\subsection{Real-Time Candidates}
\label{ssec:realtime-candidates}

\begin{figure*}
    \centering
    \hspace{0.8cm}
    \begin{subfigure}{}
      \centering
      \includegraphics[width=.46\linewidth]{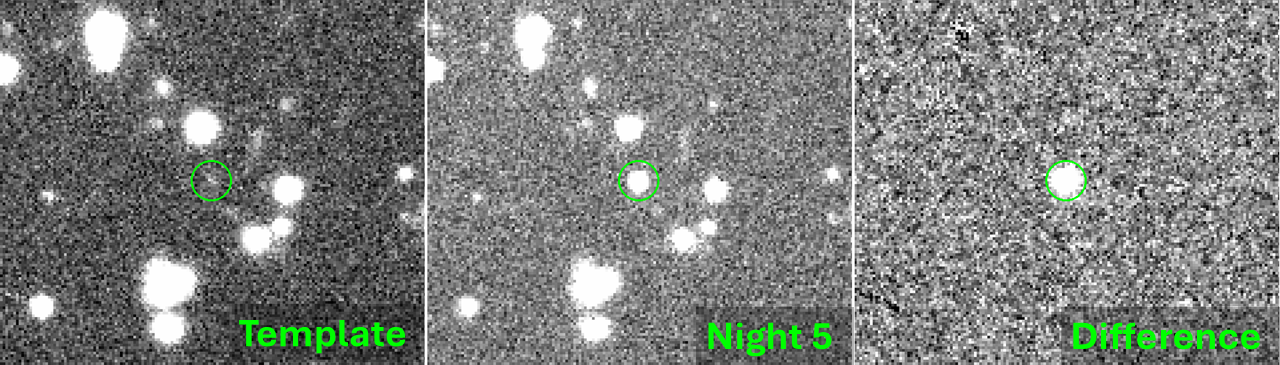}
    \end{subfigure}%
    \begin{subfigure}{}
      \centering
      \includegraphics[width=.46\linewidth]{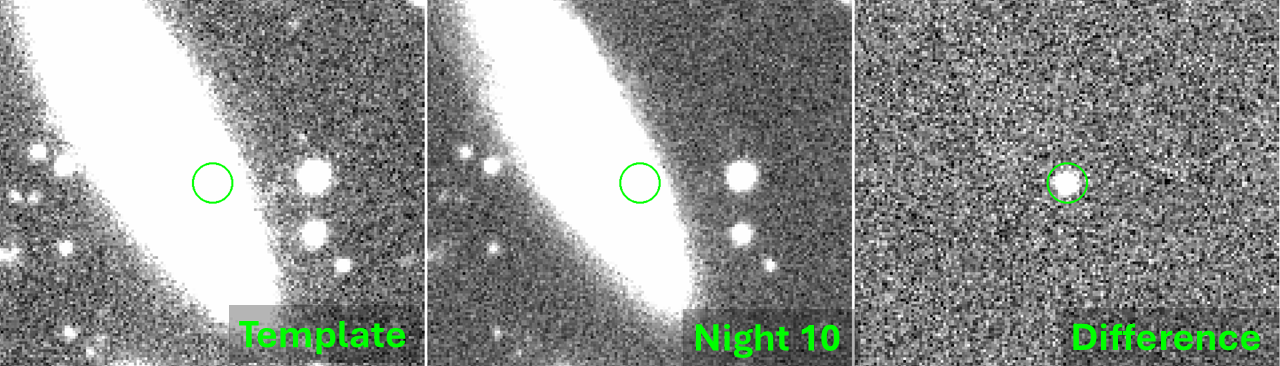}
    \end{subfigure}%
    
    \includegraphics[width=1\linewidth]{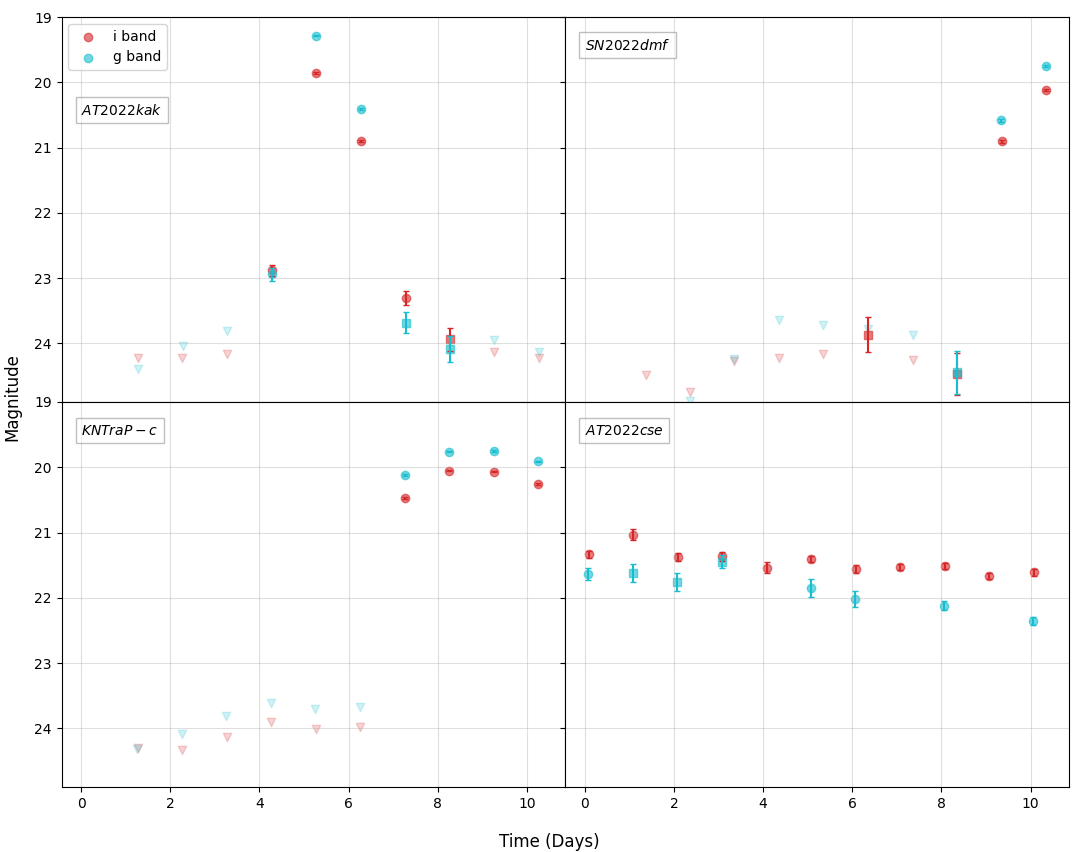}

    \hspace{0.8cm}
     \begin{subfigure}{}
      \centering
      \includegraphics[width=.46\linewidth]{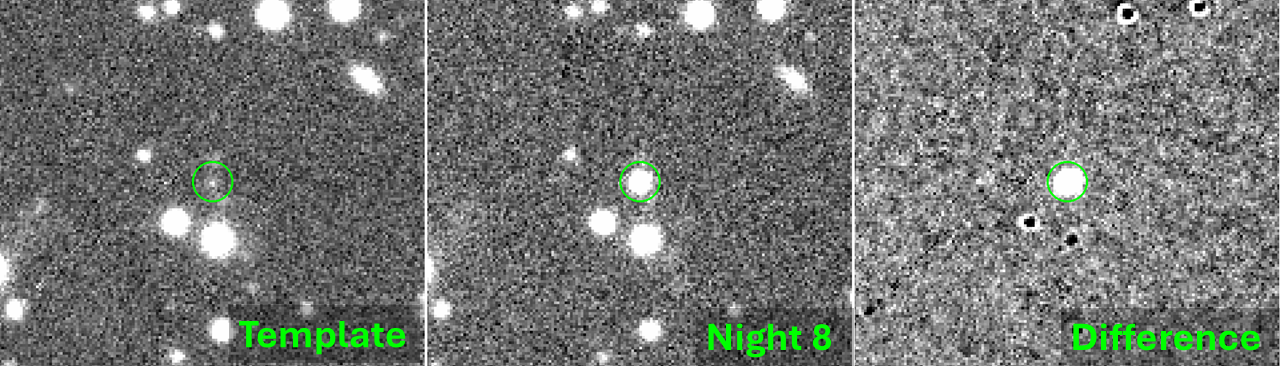}
    \end{subfigure}%
    \begin{subfigure}{}
      \centering
      \includegraphics[width=.46\linewidth]{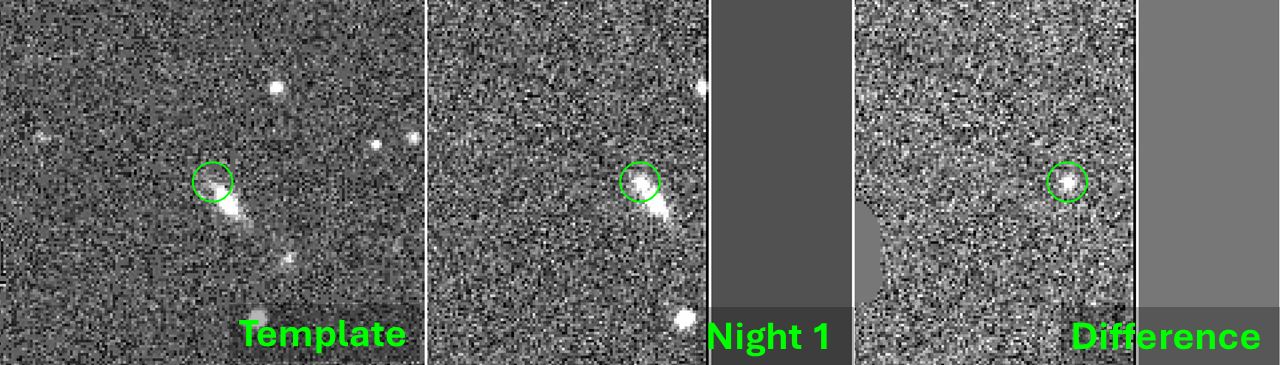}
    \end{subfigure}%
    
    \caption{Light curves of the real-time candidates in central panel described in \ref{ssec:realtime-candidates}, shown in \emph{i} (red) and \emph{g} (blue) filters. Clockwise from top-left, are the candidates \emph{AT2022kak}, \emph{SN2022dmf}, \emph{AT2022cse}, and \emph{KNTraP-c}. Panels above and below the central panel show thumbnail images for each candidate in \emph{i}-band, from left to right the images are the template, science, and difference. Night 0 was used as the template for all candidates shown here, except \emph{AT2022cse}, which used a pre-existing template. Green circles indicate the location of the transient in each image. Each circle has a radius of two arcseconds. The thumbnails of candidate \emph{AT2022cse} show a CCD edge on the right hand side of the images, but the candidate is not close enough to the edge to be rejected.}
    \label{fig:realtime-candidates}
\end{figure*}

During the observing run, using the real-time pipeline described in \S\ref{sssec:realtime_processing}, 38,530 candidates were produced from {\tt Photpipe}. There were $\sim$50 candidates identified in real-time as potentially interesting. The Neil Gehrels Swift Observatory \citep{Gehrels:2004}, the extended Public ESO Spectroscopic Survey of Transient Objects \citep[ePessto+,][]{Smartt:2015}, the Gamma-ray Burst Optical/Near-infrared Detector \citep[GROND,][]{Greiner:2008}, Zadko Observatory \citep{Coward:2017}, and the Australia Telescope Compact Array \citep[ATCA,][]{Frater:1992} were used in real-time and late-time follow-up efforts for interesting real-time candidates. Due to earlier than expected scheduling and other complications, we were not able to trigger most telescopes until later in the run. Three fast-rising/evolving real-time candidates and one interesting bright candidate are presented in the following subsections. Their photometry can be found in Appendix \ref{appsec:realtime_photometry} and their light curves and thumbnail images are shown in Figure \ref{fig:realtime-candidates}.

%\subsubsection{KNTraPaaa}
\subsubsection{AT 2022kak}
This candidate \citep[AT2022kak,][]{Zhang_kak:2022} was fast-rising and peaked on the fifth night of the observing run and was bright for two nights. It was caught on the rise and rose $\sim$3-4 mags in the first day, peaked at m($g$) = 19.23 and m($i$) = 19.78, and faded rapidly over two nights with an average fade rate of $\Delta$m($i$) = 1.72 mag day$^{-1}$. GROND was triggered on source, detecting it $\sim$3 days after the last KNTraP detection, and monitored the source for a further $\sim$1.5 months. A Swift ToO was triggered one week after the end of the KNTraP run, giving upper limits of 21.78 AB mag with the UltraViolet and Optical Telescope (UVOT) in the uvm2 filter. Swift's X-Ray Telescope (XRT) was also used to get an upper limit in the 0.3--10 keV range with a flux of 3.6 $\times10^{-14}$ erg/cm$^{2}$/s. Assuming an absorbed power law with a Galactic column density of 6.83$\times10^{20}$ cm$^{2}$, and a photon index of 2 at redshift z $\sim$ 1, the 3$\sigma$ luminosity upper limit is 3.2$\times10^{44}$ erg/s. If the GW170817 redshift were z $\sim$ 0.03, the 3$\sigma$ X-ray luminosity upper limit is 8.1$\times10^{40}$ erg/s. 

This transient was located on a persistent faint non-PSF source in our DECam images, which may be the host galaxy. This source has a magnitude of 23.16 in the \emph{i}-band, too faint to be in the online catalogs used in the candidate selection. It was also cross-matched with the DECam Local Volume Exploration Survey \citep[DELVE,][]{Drlica-Wagner:2021} without a conclusive identification. 
There are two nearby and non-PSF sources (m($i$) = 22.99 and m($i$) = 24.34 mag) which may be host galaxy candidates, also not found in online catalogs. 

If the transient was a kilonovae with a peak luminosity similar to AT 2017gfo, that would place it at a distance of $\sim$150 Mpc and the underlying source, if the host galaxy, would have a luminosity of M($i$) $\sim$ $-$12.7. The fast fading behaviour and lack of reddening makes this an interesting candidate and is investigated in a future paper (Van Bemmel et al. in prep).

%\subsubsection{KNTraPaab}
\subsubsection{SN 2022dmf}
This fast-rising candidate was observed on the final two nights of the observing run, and rose by $\sim$3 mags on the first day to 20.17 mag in the \emph{i}-band and 19.75 mag in the \emph{g}-band. The transient was located in a galaxy, but offset from the galaxy's center, making it an interesting candidate. The galaxy is identified in the SIMBAD catalog as 2MFGC 11439. GROND, Swift, and ePESSTO+ were triggered for follow-up on this source. The spectra obtained from ePESSTO+ confirmed that this source is a Type Ia supernova, and given the identifier, SN2022dmf \citep{Smith_dmf:2022}. This supernova was also independently identified by ATLAS. As there were only two consecutive data points seen for this candidate, it did not pass the offline KN vetting process. The early detection of this faint Type Ia is however interesting, and further analysis may help constrain the progenitor of this event.

%\subsubsection{KNTraPaac}
\subsubsection{KNTraP-c}
This candidate appeared the eighth night of observing and was visible for four nights until the end of the run. It rose $\sim$4 mags on the first day and reached peak magnitude m($g$) = 19.77 and m($i$) = 20.05. GROND was used to gather follow-up information, and was able to observe this source a week after the last KNTraP detection, and monitored for one month . The object location has been crossmatched with a faint source in DELVE, with magnitude m($g$) = 24.05 and m($i$) = 22.36. This candidate was rejected from the offline KN filtering as its evolution was too slow, fading 0.189 mags between the final two data points in the \emph{i}-band. The astrophysical origin of this transient remains unknown. Further detailed analysis is outside the scope of this paper, as it does not meet the KN criteria in this work.

%\subsubsection{KNTraPaad}
\subsubsection{AT 2022cse}
This candidate was seen through the entire observing run, and is included here as an interesting real-time transient. It did not survive the KN filtering, as it was visible on the first night, and faded too slowly to be considered a KN candidate, fading on average 0.06 mag day$^{-1}$ in the \emph{i}-band, and 0.07 mag day$^{-1}$ in the \emph{g}-band. The source is located on the edge of a galaxy. There was no cross-match of this source in Gaia, SIMBAD, or Vizier. In real time, this source was also identified and classified as a likely supernova candidate, and given the designation AT2022cse \citep{Zhang_cse:2022}. The likely host galaxy is found in the DESI Legacy Survey with a phot-z of 0.37$\pm$0.073. Using this redshift and assuming the peak in our data reflects the overall peak of the transient, its peak absolute magnitude is estimated at M(\emph{i}) $\sim$ $-$20.21$^{+0.39}_{-0.41}$. This luminosity corresponds to the high-end tip of the luminosity distributions for supernovae \citep{richardson14}.

\begin{figure*}
    \centering
    \includegraphics[width=1\linewidth]{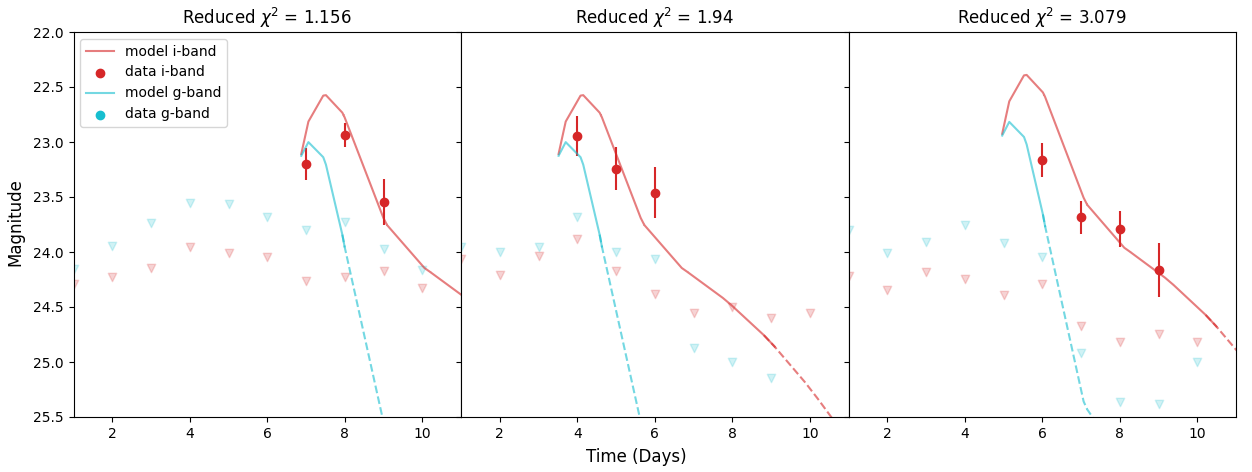}
    \caption{The best fitting light curves to a KN model. The model is shown as a solid line in both \emph{i} and \emph{g}-bands above the limiting magnitude, and as a dashed line below the limiting magnitude. The candidate data points shown as circles. Limiting magnitudes are shown as inverted triangles on each night as a reference. The thumbnail images of these sources reveal they are not real transients, however these plots highlight how artefacts can mimic KN light curve signatures.}
    \label{fig:chi2_bestfits}
\end{figure*}

\subsection{Offline Candidates}
\label{ssec:offline_candidates}

After the consecutive detection candidate selection cut described in Section \S\ref{sssec:consecutive_detections}, there were 1436 candidates remaining, as seen in more detail in Table \ref{tab:candsummary}. These candidates then underwent the $\chi^{2}$ model fitting search, and candidates fitted against a KN model resulting in a p-value of $>$ 0.001 were selected. There were 52 matches to this criteria, which were inspected manually using their light curves and thumbnail images.

The best fitting candidate has a p-value of 0.328 and reduced $\chi^2$ value of 1.156. The light curve of this candidate is shown in  the left panel in Figure \ref{fig:chi2_bestfits}. Assessing the thumbnail images of this candidate reveals that it is the product of a bad subtraction, and only coincidentally follows the shape of the KN model. Viewing all of the thumbnails of the light curves, there are no images which support any of these light curves being real transient objects. A majority of these light curves appear on the seventh observing night, which is when the seeing and lunation begins to improve. As these light curves are near the limiting magnitude, it is possible that due to improved observing conditions, these faint object subtraction residuals just begin to peak over our detection threshold on those nights. 

There were 776 candidates which passed through the rapid evolution vetting. With the bad subtractions, artefacts, variable, and steady sources removed, there remained 11 fast fading candidates. Doing further searches into these candidates, all were found to be cross-matched with Gaia. These candidates were not removed in prior filtering, as their parallax values were below the threshold for stellarity. All but one were classified with high confidence as being a single star, and six flagged as variable, the remaining one was classified as a quasar and present in the Gaia quasar catalog. 

\begin{figure}
    \centering
        \scalebox{0.57}[0.58]{\includegraphics{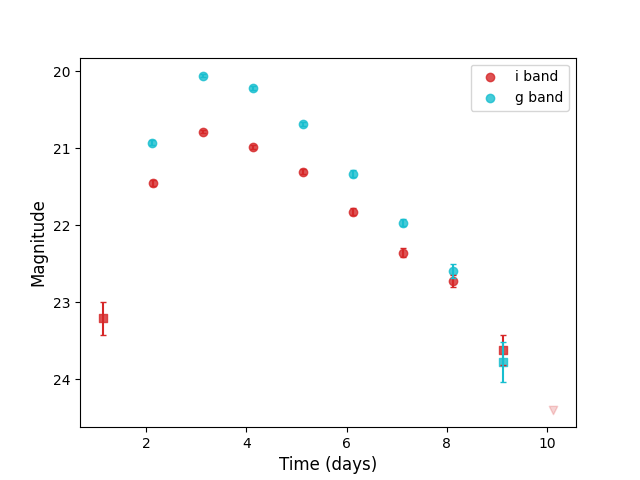}}
        \caption{An RR Lyrae variable star, which when sampled nightly as with KNTraP, has the appearance of a fading transient object in both filters. Circle markers indicate unforced photometry, and square markers indicate forced photometry. Limiting magnitudes are shows as inverted triangles.}
        \label{fig:variable_contaminant}
\end{figure}

A common contaminant identified were variable stars (e.g., RR Lyrae and eclipsing binaries), of which $\sim$200 passed through the rapid evolution filtering. Some of these variable objects appear to have light curves similar to fast fading transients when observed near the same time each night. These variable stars are expected to have hours-long periods, but due to a sampling effect caused by our consistent nightly cadence, we are seeing longer-scale variability. An example of this is shown in Figure \ref{fig:variable_contaminant}. Although variable sources should be eliminated by our selection process because they are present on the first night, variable sources in fields with template images at a phase in their evolution that had light curve flux that matched the template flux are not detected in the first night's detection image. Also, many catalogs do not cover the faint variables present in the depths reached by this survey.

No convincing candidates that met our criteria were found in this search. It was expected that the likelihood of a KNe meeting our criteria would be low, as the initial expected KN discovery rate for this run was $\sim$0.3 KN run$^{-1}$, the observing conditions were not ideal (bright/grey lunation) and the proximity of the Galactic plane) led to the magnitude depths reduced by $\sim$1 mag. The strategy is for more than 3+ runs to detect KNe.

To test the real-time candidates, and to validate the selection process, they were run through the KN offline filtering; a summary of the results is shown in Table \ref{tab:candsummary}. Of the 50 real-time candidates, 23 were classified as non-variable objects. None of these 23 candidates pass all the selection criteria. Ten were removed as they were detected on the first night. A further nine were rejected as they did not have at least three consecutive PSF-like detections in the \emph{i}-band. Two promising candidates of these nine were removed this way as they only had two consecutive PSF-like detections. One of these was rising on the last two nights of the run, and though it was removed through this filtering, it remained to be monitored in real-time by follow-up instruments. Finally, the remaining four real-time candidates were rejected as their evolution was too slow to be KN-like, or were only rising.

\begin{figure*}
    \centering
    \includegraphics[width=1\linewidth]{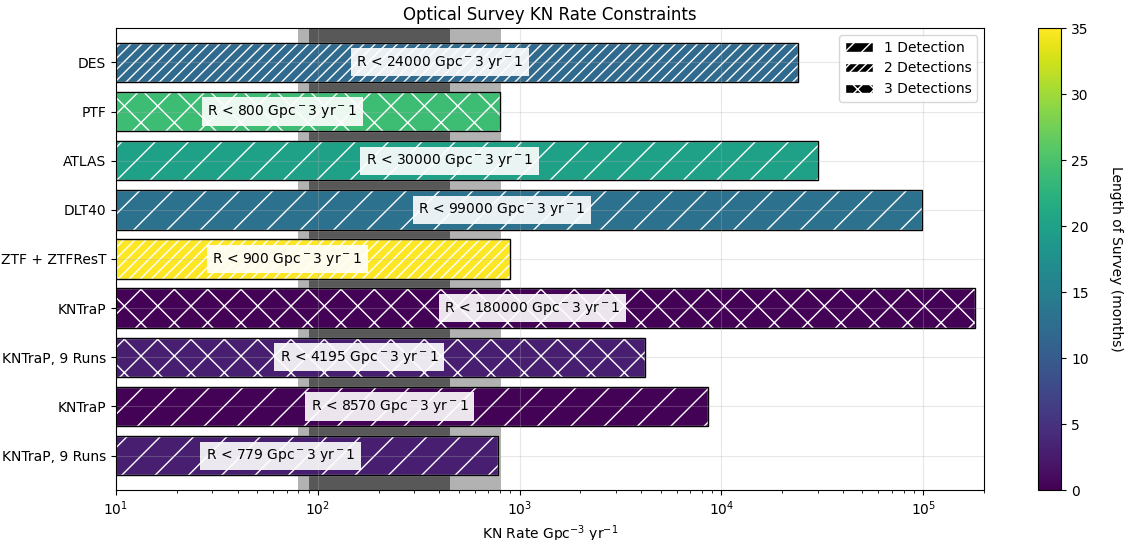}
    \caption{Comparisons between the KN rate constraints made by optical surveys \citep{Doctor:2017, Kasliwal:2017, Smartt:2017, Yang:2017, Andreoni:2021}, and the upper limits calculated by KNTraP. The number of detections required for a KN detection for each survey us shown with hatching, and the length of the survey in months is indicated with colour. The light grey shaded region shows the intrinsic BNS merger rate \emph{$R_{BNS} = 320_{-240}^{+490}$} Gpc$^{-3}$ yr$^{-1}$ \citep{Abbott:2021}, and the dark grey region shows the most current BNS merger rate constraints, \emph{$R_{BNS} = 210_{-120}^{+240}$} Gpc$^{-3}$ yr$^{-1}$ \citep{theligo-virgo-kagracollaboration_observing_2023}. KNTraP rates shown are for 50\% recovery rates, with one and three detections required for KN candidates, for both the first run reported in this work as well as the projected limits after nine KNTraP runs, assuming optimal conditions at a 95\% confidence upper limit.}
    \label{fig:rates_comparison}
\end{figure*}

\section{Kilonova Rates}
\label{sec:rates}
This survey using a consistent nightly cadence in two filters provides a constraint on the optical rate on AT2017gfo-like KNe that are detected beyond the LVK horizon and is not reliant on GW detections or compact object merger orientations. We used the KN model grid discussed in \S\ref{sec:kn_models} to simulate KNe at different redshifts and starting dates. The recovery rate of these simulated AT2017gfo-like KNe with our survey strategy informs us of our detection and selection efficiency.

Using the conservative filtering methods described above, the distance that 50\% (90\%) of AT2017gfo-like KNe with at least three $\geq$3$\sigma$ photometric data points (typically peaking $\sim$1 mag above the magnitude limit) in the \emph{i}-band could be recovered using the described filtering methods is $z \sim 0.15$, or 650 Mpc ($z \sim 0.13$ or 570 Mpc). These distances give nightly volumes of $\sim2.3\times10^{-3}$ Gpc$^{3}$ and $\sim1.4\times10^{-3}$ Gpc$^{3}$ for the 50\% and 90\% recovery rates, respectively. Galactic extinction corrections were made per filter for the magnitude limits before calculating the volumes. Given our 10-night observing window (first night used as the template), and that we required  a conservative non-detection on the first science night, the first two nights are not used for KN detection, reducing the observing window for the rate calculation from 11 to 9 nights.
In addition, as we require at least three consecutive data points for a KN detection, the final two observing nights are not included in the rate calculation, as any KN occurring during those two nights would not be considered here. 

With our conservative criteria, there remains a seven night observing window to capture and classify KNe. More potential candidates could be found if these cuts were to be relaxed, however we only want to consider transients which could be considered strong AT2017gfo-like candidates and those that meet our rapid evolution criteria.

We compute a rate limit where three AT2017gfo-like KNe could be detected for a 95\% confidence level, as informed by small number Poissonian statistics  \citep{Gehrels:1986}, given no compelling KN candidates met our selection criteria during this run. For 50\% and 90\% KN recovery rates from the model grid, we find the rate upper limit to be $R < 1.8\times10^{5}$ Gpc$^{-3}$ yr$^{-1}$ and $R < 2.6\times10^{5}$ Gpc$^{-3}$ yr$^{-1}$. In general, this rate is not comparatively very constraining at this point in the KNTraP program, but Figure \ref{fig:rates_comparison} shows our rate is consistent with the KN rate limits as calculated by other optical surveys, which have all used substantially larger amounts of time on sky. If we consider lowering our threshold for a KN classification from requiring three detections to only a single detection, as is done with the ATLAS and DLT40 surveys, our rate constraint becomes $R <$ 8570 Gpc$^{-3}$ yr$^{-1}$ (for 50\% KN recovery).

\section{Discussion}
\label{sec:discussion}
The KNTraP strategy is to probe a large volume of the sky, with multiple filters, using a balance between area on sky and depth afforded by DECam. The design is an efficient search for KNe, with the ability to search beyond the LVK horizon, search for redshift evolution, discover KNe without reliance on the inclination angle, avoid interruptive searches, and to catch KNe before outburst and during their rise, peak, and fade.  

This first KNTraP run incurred some limitations. Due to the lunation on our scheduled nights, the magnitude limits were $\sim$1 mag shallower than the optimal program aims. In addition, the high visibility of the Galactic plane at the time of scheduling meant that fields with somewhat higher than optimal Galactic extinction had to be chosen. For a KNTraP run in dark time, low Galactic extinction, and similar magnitude depths, about twice as many fields can be observed due to the lower amount of exposure time needed. This increased area on sky leads to a $\sim$2 times improvement in nightly volume probed, and a rate $\sim$6 times more constrained ($\sim R < 3.1\times10^{4}$ Gpc$^{-3}$ yr$^{-1}$, for 50\% recovery). Figure \ref{fig:volume_limits}, shows that the strategy selected in this run was optimal for the given conditions. 

In addition, the depths reached for our offline search was reduced to $z \sim$ 0.15 due to the constraint of requiring at least three detections for a KN candidate, compared to $z \sim$ 0.3 using a minimum of one detection. Future KNTraP runs of the same fields will be more efficient, as the data gathered from this run can be used to provide template images, removing the need to create template images during the run and requiring a first science night non-detection. This allows for KNe rising on the first day, and for two more days to probe KN for their rate. 

Nevertheless, this observing run was a successful proof of concept. Using the intrinsic BNS rate calculated by \cite{Abbott:2021}, we need three KNTraP runs find one KN on average, and up to nine runs to find one KN to a 95\% confidence limit (i.e., computed to detect 3 KNe) with the consideration of small number statistics. This estimate is using our conservative cuts and requirements to define a KN detection. The expected rate of KN calculated for this survey is dependent upon the intrinsic BNS rate, which is subject to change with further GW observational runs, and does not factor in short and long GRB rates, which KNe candidates have been association with \citep{Perley:2009, Tanvir:2013, Berger:2013, Gao:2015, Jin:2015, Jin:2016, Jin:2020, Troja:2018, Lamb:2019, Rastinejad:2022, Levan:2023}. 

This run also proved that interesting fast-evolving transients can be found to large distances alongside the search for KNe when using nightly cadence and two-colour evolution. In real-time, several fast-rising transients were identified and followed-up with other instruments and determined to be non-KN (e.g., mostly supernovae and other catalogued bursting and variable events).

The estimated KN rate upper limit from this first KNTraP run is in agreement, but is not yet more constraining than other optical surveys. A comparison of KN rates from several optical surveys are shown in Figure \ref{fig:rates_comparison}, which details the lengths of the surveys and the number of detections required to ascertain a KN. The rates calculated with ATLAS, DLT40, and ZTF explicitly state that Poisson small number statistics are used to ascertain a 95\% confidence upper limit. The DES and PTF confidence limits are 90\% and 99.7\% (3$\sigma$) respectively. 

One of the KN upper limits was calculated using two years of data from the Dark Energy Survey supernova program \citep[DES-SN,][]{Bernstein:2012,  Diehl:2016}, which also used DECam. For the case of a bright KN (this survey was conducted before AT2017gfo), the upper limit was $R < 2.4\times10^{4}$ Gpc$^{-3}$ yr$^{-1}$ \citep{Doctor:2017}. The DES search reached similar magnitude depths to KNTraP, with multiple filters, however had a one week cadence - therefore a KN found this way would likely only have a single detection per filter and less chance of catching a KN fast evolution. With the same instrument, KNTraP constrains the KN rate in agreement to the rate constrained by DES, but using much less time, and with stricter requirements on what is considered a KN detection. With similar constraints on the selection criteria for a KN candidate (requiring a single detection), KNTraP is estimated to constrain the KN rate to be $R < 8570$ Gpc$^{-3}$ yr$^{-1}$ for this run. 
 
Current pathways to KN discovery largely rely on GW and GRB triggers, which in the past two LVK observing runs, O3 and O4, have seen many triggers for follow-up observations by various telescopes. From what can be gathered from the published literature, there has been GW event follow up in O3 on 15 nights, which totalled to 31-51 hours on DECam, with no KN detections. This is not inclusive of the hours spent on follow-up with DECam which has not been published in O3 and does not include GW follow up time in O4. In addition, there are other optical telescopes which in O3 and O4 are triggered for follow-up observations, as well as efforts for GRB optical follow-up, resulting in a large cost in personnel and telescope follow-up time, with many efforts resulting in disruptive time. In comparison, the KNTraP strategy requires $\sim$30 nights (270 hours) per KN detection using DECam on classically scheduled nights. 

It may be that conventional follow-up is similarly effective in detecting KNe, however, KNTraP is non-interruptive to telescope time, is predictable to arrange follow-up spectroscopy and imaging, reaches well beyond the GW detector horizons, and obtains the event (and other transients) before, during, and after the outburst. Moreover, KNTraP can optimise field selection to maximise the chances of detection, avoiding the Galactic plane and the Sun, which can be issues with GW/GRB follow-up. Conventional GW follow-up also means that the early stages of the optical light curve can be missed. However, as GW data is stored, a search/subthreshold search can be employed for any KN candidate found by an optically led search. Similarly, an archival subthreshold search for coincident GRB signals can be conducted. However, given the DECam FOV and rate of GRB and GW events, a coincident subthreshold detection may be unlikely, so follow-up of such events remains the main avenue.

There exist other optical surveys such as ZTF, which can conduct searches similar to KNTraP. Due to the fast evolution of KNe, a nightly cadence with two filters per day is preferred. ZTF is now conducting pilot surveys which has fast cadence and multi-filter capabilities \citep{Ahumada:2023, Ho:2024}, and selection mechanisms for KN candidates have been implemented \citep{Andreoni:2021, Biswas:2023}. However, none of the followed-up ZTF candidates have been confirmed to be a KN \citep{Andreoni:2021}, nor through searches using the ZTF public stream by {\tt Fink} and GRANDMA \citep{Moller:2021, Aivazyan:2022}. Shallower optical surveys can make up comparative volume through large area on sky over longer periods of time, but are not able to reach beyond the GW detector horizons. With the low expected rate of KNe, dedicated KN searches with a smaller optical telescopes take much more time, and hence more personnel effort and cost. In addition, follow-up over these longer surveys need to be fully triggered, as scheduling follow-up over many months is difficult, however over a shorter time-span ($\sim$1 week), this is more feasible.

Upcoming instruments and surveys such as the Legacy Survey of Space and Time (LSST) conducted by the Vera C. Rubin Observatory \citep{Ivezić:2019} and surveys with the Nancy Grace Roman Space Telescope \citep{Spergel:2015} can search for KNe. However, their main baseline surveys will not have nightly cadence, greatly limiting the ability and confidence of any KN candidate selection. Although $>$300 KNe can be discovered out to $\sim$1400 Mpc with Rubin over the 10-year LSST survey, only 3--32 will be recognisable as AT2017gfo-like transients \citep{Andreoni_RubinKN:2022, Hambleton:2023}. Recommendations have been made for nightly multi-filter observations \citep[e.g.][]{Andreoni:2019, Bianco:2019, Setzer:2019}, and for simulations to explore a ``rolling cadence'' strategy which allows fields to be observed multiple times per night in multiple filters \citep{Andreoni_RubinKN:2022}. \cite{Setzer:2019} showed that with a more optimised strategy for KNe (within a night, a field must have repeat observations with different filters), the expected number of KN detections can be improved from the current baseline strategy by at least three times. A KNTraP-like survey designed for a more sensitive instrument such as the Keck Wide-Field Imager \citep[KWFI,][]{Radovan:2022, Radovan:2024}, would improve efficiency. This improvement is demonstrated in Figure \ref{fig:volume_limits}, where a larger nightly volume can be probed with KWFI versus DECam for a KNTraP survey, detecting KNe with $\sim$3 times less observing time. The optimal KWFI strategy would reach $\sim$25 mag, and target 145 fields ($\sim$55 equivalent DECam sized fields) per night, observing a volume of $\sim$0.05 Gpc$^{3}$ nightly.

\section{Summary}
\label{sec:summary}
The aim of KNTraP was to optimise an optical search to find KNe without GW or GRB triggers. KNTraP used DECam as it is the optimal instrument for this search due to its sensitivity, wide FOV, and fast filter exchange. This strategy allows us to probe volumes of the Universe, past the GW detector horizon.

The data were searched using a rapid real-time pipeline and a rigorous offline pipeline. Despite the less than optimal observing conditions, early-detected and fast-evolving transients were found. There were four real-time candidates determined to be very interesting, however further analysis showed two candidates to be evolving too slowly, and remained too blue to be KN-like. One candidate was spectroscopically confirmed to be an early Type Ia, and the last candidate remains an object of interest, although it is not AT2017gfo-like. This candidate will be explored further in a future paper.

Through rigorous and conservative offline filtering after the run, $\sim3\times10^{5}$ candidates identified through {\tt Photpipe} were reduced to $\sim$1000 with the requirement not be a known stellar object, have at least three detections in any filter, not be present on the first science night, and not be near the edge of the CCD. To identify convincing KN candidates, they were required to have a minimum of three consecutive detections in a single filter that are PSF-like and at least 1$\sigma$ above the limiting magnitude. 

Two methods were used to test these candidates: (1) a $\chi^{2}$ fitting method was used to test the candidates against a model best fit to AT2017gfo by \cite{Villar:2017} and (2) by vetting out the non-rapidly fading candidates and inspecting the remaining candidates by eye. Through these two methods, no convincing AT2017gfo-like candidates were found. This result is not surprising, as we estimate $\sim$ 0.3 KNe with a single KNTraP run using the BNS merger rate of \emph{$R_{BNS} = 320_{-240}^{+490}$} Gpc$^{-3}$ yr$^{-1}$ \citep{Abbott:2021}. This means on average, the expectation is that it would take a further two KNTraP runs to find one KN and up to eight additional runs to achieve a 95\% confidence limit using small number statistics. 

Results from the survey allows us to estimate a 95\% confidence upper limit to the KN rate to be $R < 1.8\times10^{5}$ Gpc$^{-3}$ yr$^{-1}$ and $R < 2.6\times10^{5}$ yr$^{-1}$ for 90\% and 50\% recovery rates of a distribution of AT2017gfo-like light curves. 

The design of KNTraP has advantages over conventional GW and GRB follow-up as it can reach beyond the GW detector horizon distances, search for redshift evolution, search fields with low Galactic extinction and away from the Sun and Galactic plan (in which some GW search areas fall), and the search is not sensitive to merger system orientations. An optically led search such as KNTraP allows for the random sampling of very early stages of KN light curves, whereas conventional follow-up methods could miss this via the nature of their reactive response. In addition, KNTraP is non-interruptive and, as it is classically-scheduled, is predictable to organise follow-up spectroscopy and imaging. GW follow-up discovery efficiency is highly dependent on field pointings and search area, and KNTraP is comparable to a GW direct follow-up search if the search area is in the hundreds to a thousand square degree range. Future KNTraP runs are encouraged to help accelerate KN discovery, and would benefit from using this pilot run as a template. Future instruments with higher sensitivities can also be utilised for even higher efficiency searches.

\section*{Acknowledgements}
This research was funded in part by the Australian Research Council Centre of Excellence for Gravitational Wave Discovery (OzGrav), CE170100004 and CE230100016. 
JC acknowledges funding from the Australian Research Council Discovery Project DP200102102. 
AM is supported by DE230100055. 
IA acknowledges Amazon Web Services. 
D.O.J. acknowledges support from HST grants HST-GO-17128.028 and HST-GO-16269.012, awarded by the Space Telescope Science Institute (STScI), which is operated by the Association of Universities for Research in Astronomy, Inc., for NASA, under contract NAS5-26555. D.O.J. also acknowledges support from NSF grant AST-2407632 and NASA grant 80NSSC24M0023.
Part of the funding for GROND (both hardware as well as personnel) was generously granted from the Leibniz-Prize to Prof. G. Hasinger (DFG grant HA 1850/28-1).

%%%%%%%%%%%%%%%%%%%%%%%%%%%%%%%%%%%%%%%%%%%%%%%%%%
\section*{Data Availability}

Calibrated DECam images from this survey are publicly available on the NOIRLab Astro Data Archive under program number 2022A-679480. The web portal can be viewed at \url{https://astroarchive.noirlab.edu}.

% The inclusion of a Data Availability Statement is a requirement for articles published in MNRAS. Data Availability Statements provide a standardised format for readers to understand the availability of data underlying the research results described in the article. The statement may refer to original data generated in the course of the study or to third-party data analysed in the article. The statement should describe and provide means of access, where possible, by linking to the data or providing the required accession numbers for the relevant databases or DOIs.



\bibliographystyle{mnras}
\bibliography{references} % if your bibtex file is called example.bib

%\begin{thebibliography}{99}
%\end{thebibliography}

%%%%%%%%%%%%%%%%%%%%%%%%%%%%%%%%%%%%%%%%%%%%%%%%%%

%%%%%%%%%%%%%%%%% APPENDICES %%%%%%%%%%%%%%%%%%%%%

\appendix
%If you want to present additional material which would interrupt the flow of the main paper, it can be placed in an Appendix which appears after the list of references.

\section{AT2017gfo Model Grid}
\label{appsec:model_grid}

\begin{figure*}
    \centering
    \includegraphics[width=1\linewidth]{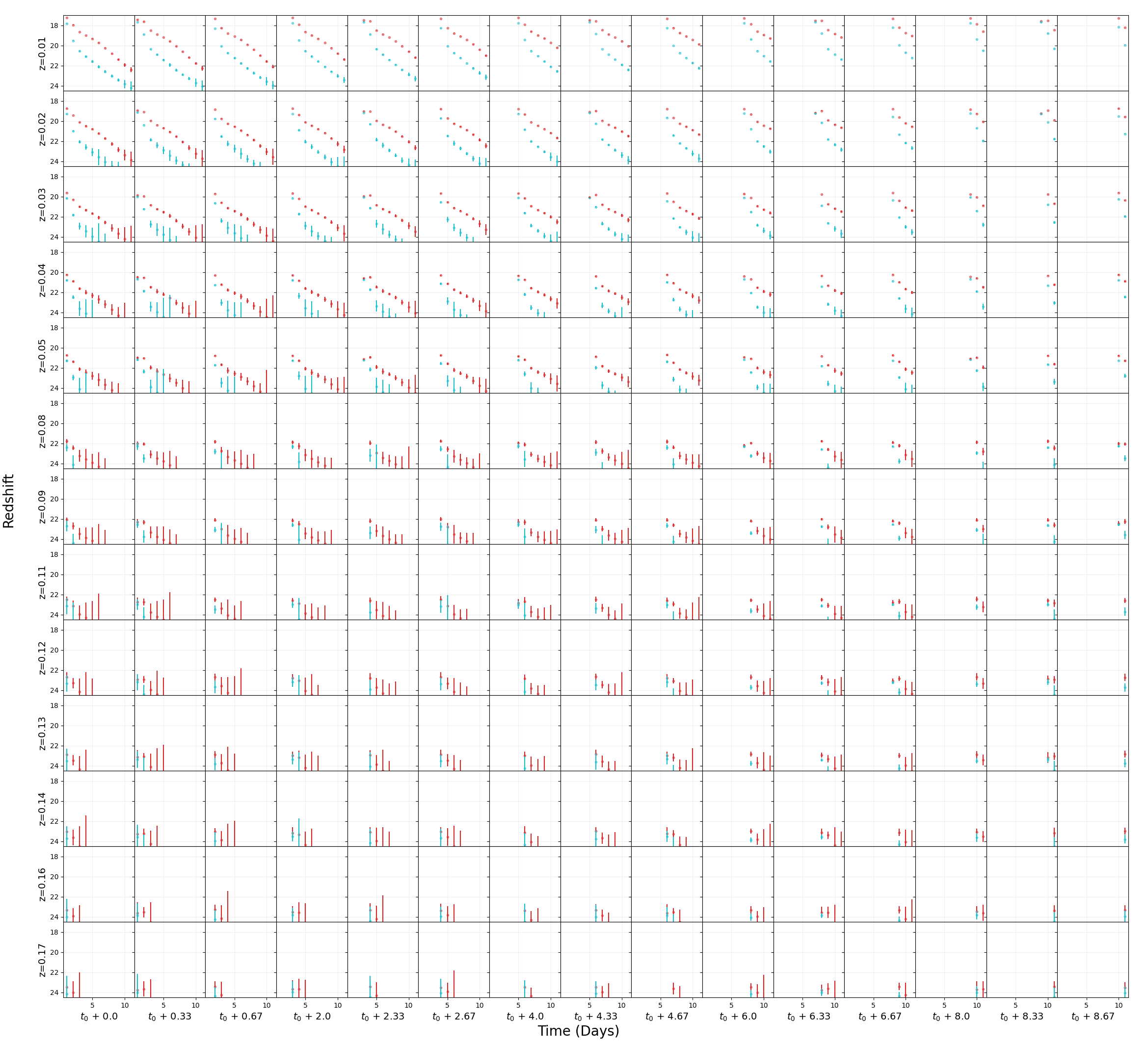}
    \caption{Representative subset of the KN model grid as described in Section \S\ref{sec:kn_models}. Light curves shown in \emph{i}-band (red) and \emph{g}-band (blue). KN models increase in redshift down the y-axis, and burst date $t_{0}$ is shifted along the x-axis.}
    \label{fig:model_grid}
\end{figure*}

In Figure \ref{fig:model_grid} we present the grid of AT2017gfo model light curves.

\section{Real-Time Candidates Photometry Table}
\label{appsec:realtime_photometry}

In Table \ref{tab:realtime_candidate_photometry} we present the photometry of the four most interesting fast transient candidates found in real-time.

\begin{table}
    \centering
    \caption{Photometry table for real-time candidates.}
    \begin{tabular}{|c|c|c|c|}
        \hline
        MJD & Filter & Magnitude & Magnitude Error \\
        \hline
        \multicolumn{4}{|c|}{\emph{KNTraPaaa}} \\
        \hline
        59626.275 & g & 22.942 & 0.101 \\
        59626.273 & i & 22.879 & 0.085 \\
        59627.272 & g & 19.286 & 0.010 \\
        59627.270 & i & 19.858 & 0.011 \\
        59628.273 & g & 20.413 & 0.012 \\
        59628.270 & i & 20.893 & 0.015 \\
        59629.280 & i & 23.306 & 0.102 \\
        \hline
        \multicolumn{4}{|c|}{\emph{KNTraPaab}} \\
        \hline
        59631.348 & g & 20.585 & 0.022 \\
        59631.350 & i & 20.902 & 0.023 \\
        59632.348 & g & 19.754 & 0.012 \\
        59632.350 & i & 20.117 & 0.013 \\
        \hline
        \multicolumn{4}{|c|}{\emph{KNTraPaac}} \\
        \hline
        59629.267 & g & 20.116 & 0.012 \\
        59629.269 & i & 20.470 & 0.013 \\
        59630.262 & g & 19.759 & 0.011 \\
        59630.264 & i & 20.053 & 0.011 \\
        59631.255 & g & 19.753 & 0.011 \\
        59631.257 & i & 20.066 & 0.012 \\
        59632.253 & g & 19.910 & 0.011 \\
        59632.255 & i & 20.255 & 0.012 \\
        \hline
        \multicolumn{4}{|c|}{\emph{KNTraPaad}} \\
        \hline
        59622.071 & g & 21.637 & 0.090 \\
        59622.080 & i & 21.334 & 0.055 \\
        59623.083 & i & 21.031 & 0.085 \\
        59624.081 & i & 21.371 & 0.060 \\
        59625.081 & i & 21.362 & 0.068 \\
        59626.081 & i & 21.539 & 0.082 \\
        59627.070 & g & 21.849 & 0.134 \\
        59627.078 & i & 21.404 & 0.051 \\
        59628.069 & g & 22.016 & 0.126 \\
        59628.078 & i & 21.554 & 0.062 \\
        59629.079 & i & 21.531 & 0.049 \\
        59630.068 & g & 22.120 & 0.074 \\
        59630.077 & i & 21.514 & 0.045 \\
        59631.076 & i & 21.672 & 0.047 \\
        59632.065 & g & 22.356 & 0.056 \\
        59632.074 & i & 21.610 & 0.056 \\
        \hline
    \end{tabular}
    
    \label{tab:realtime_candidate_photometry}
\end{table}

%%%%%%%%%%%%%%%%%%%%%%%%%%%%%%%%%%%%%%%%%%%%%%%%%%

% Don't change these lines
\bsp	% typesetting comment
\label{lastpage}
\end{document}